\documentclass[12pt]{article}

\usepackage{graphicx}

\begin{document}

\title{\Large \bf Critical review of deeply bound kaonic nuclear states}
\author{\large V.K. Magas$^1$,  E. Oset$^2$, A. Ramos$^1$, H. Toki$^3$ \bigskip \\
{\it  $^1$~Departament d'Estructura i Constituents de la Mat\`eria}\\
{\it  Universitat de Barcelona,  Diagonal 647, 08028 Barcelona, Spain} \\  
{\it $^2$~ Departamento de F\'{\i}sica Te\'orica and }\\
{\it IFIC Centro Mixto Universidad de Valencia-CSIC}\\
 {\it Institutos de Investigaci\'on de Paterna,}\\
{\it  Apdo. correos 22085, 46071, Valencia, Spain} \\ 
{\it $^3$~Research Center for Nuclear Physics, Osaka University}\\
{\it Ibaraki, Osaka 567-0047, Japan}}

\maketitle

{\large

\begin{center}
{\bf Abstract}\\
\medskip
We critically revise the recent claims of narrow deeply bound kaonic states and show that at present there is no 
convincing experimental evidence for their
existence. In particular, we discuss in details the claim of $K^-pp$ 
deeply bound state associated to a peak seen in
the $\Lambda p$ invariant mass spectrum from $K^-$ nuclear absorption reactions
by the FINUDA collaboration. 
An explicit theoretical simulation 
shows that the peak is
simply generated from a
two-nucleon absorption process, like $K^- p p  \to \Lambda p$, followed by
final-state interactions of the produced particles with the residual nucleus.
\end{center}

Over the last years the possible existence of the deeply bound states of ${\bar K}$ in the system of nucleons 
was strongly discussed in the literature. 
Phenomenological fits to kaonic
atoms fed the idea because a solution where antikaons would feel
strongly attractive potentials of the order of $-200$ MeV
 at the center of the nucleus
\cite{friedman-gal} was preferred. However, a deeper understanding of the
antikaon optical potential demands it to be linked to the
elementary ${\bar K}N$ scattering amplitude which is dominated by
the presence of a resonance, the $\Lambda(1405)$, located only 27
MeV below threshold. This makes the problem  a highly non-perturbative
one. In recent years, the scattering of ${\bar K}$ mesons with nucleons 
has been treated within the context of chiral unitary methods \cite{chiral}.
The explicit incorporation of medium effects, such as 
Pauli blocking, were shown
to be important \cite{koch} and it was soon realized that
the influence of the resonance demanded the in-medium amplitudes
to be evaluated self-consistently \cite{lutz}. 
The resulting antikaon optical potentials
were quite shallower than the
phenomenological one, with depths between $-70$ and $-40$ MeV \cite{ramos},
but could reproduce equally well the
kaonic atom data \cite{atoms}.

More recently, variational calculations of few body systems using
a simplified phenomenological ${\bar K}N$ interaction predicted
extremely deep kaonic states in $^3$He and $^4$He, reaching
densities of up to ten times normal nuclear density 
\cite{akaishi02,akaishi05}. Motivated by this finding, experiment KEK-E471
 used the
$^4$He(stopped $K^-,p$) reaction and reported \cite{suzuki-kek}
 a structure in the proton momentum
spectrum, which was denoted as the tribaryon $S^0(3115)$ with strangeness
$S=-1$. If interpreted as a $(K^-pnn)$ bound state, it would
have a binding energy of 194 MeV. However, in a recent work
\cite{oset-toki} strong criticisms to the theoretical
claims of Refs.~\cite{akaishi02,akaishi05} have been put forward, and a reinterpretation
of the KEK proton peak has been given in terms of two nucleon absorption
processes, $K^- p n \to \Sigma^- p, K^- p p \to
\Sigma^0(\Lambda) p$, where the rest of the nucleus acts as a
spectator.  Similarly the peak in 
FINUDA proton momentum spectrum from $K^-$ absorption on
$^6$Li \cite{finuda_np06} can be interpreted with
the two-nucleon absorption mechanism advocated in
Ref.~\cite{oset-toki}, as it was accepted by the authors.

Another experiment of the FINUDA collaboration has measured
the invariant mass distribution of $\Lambda p$ pairs \cite{finuda}. The
spectrum shows a narrow peak at 2340 MeV, which corresponds to the same
signal seen in the
proton momentum spectrum, namely $K^-$ absorption by a two-nucleon pair leaving
the daughter nucleus in its ground state. Another wider peak is 
also seen at around
2255 MeV, which is interpreted in Ref.~\cite{finuda} 
as being a $K^- pp$ bound state with $B_{K^- pp} =
115^{+6}_{-5}({\rm stat})^{+2}_{-3}({\rm syst})$ MeV and having a
width of 
$\Gamma = 67^{+14}_{-11}({\rm stat})^{+2}_{-3}({\rm syst})$ MeV.
In a recent work \cite{magas06} we showed that this peak is 
generated from the 
interactions of the $\Lambda$ and nucleon, produced after $K^-$ absorption, with
the residual nucleus. We here present some additional results, improved by
the use of more realistic $\Lambda N$ scattering probabilities, and summarize
the present status of the field.

The reaction
$(K^-)_{\rm stopped}  A \rightarrow
\Lambda  p  A'$ proceeds by capturing a slow $K^-$ in
a high atomic orbit of the nucleus, which later cascades down till
the $K^-$ reaches a low lying orbit, from where it is finally absorbed. 
We assume
that the absorption of the $K^-$ takes
place from the lowest level where the energy shift for atoms has been
measured, or, if it is not measured, from the level where the calculated
shift \cite{atoms} falls within the measurable range.

In the case of $^6$Li, $^7$Li, $^{12}$C, $^{27}$Al and $^{51}$V (the targets
of the FINUDA experiment \cite{finuda}) and $^9$Be, $^{13}$C, $^{16}$O (to be included into the future FINUDA experiment \cite{future_finuda}) the absorption
takes place from the $2p$ orbit, except for $^{27}$Al ($3d$) and $^{51}$V ($4f$).

The width for $K^-$  absorption from $p N$ pairs in a nucleus with mass number $A$
is given, in local density approximation by
\begin{equation}
\Gamma_A  \propto  \int d^3 \vec{r} |\Psi_{K^-}(\vec{r})|^2 \rho^2 \Gamma_m 
\propto  \int d^3 \vec{r} \frac{d^3 \vec{p}_1}{(2\pi)^3} \frac{d^3 \vec{p}_2}{(2\pi)^3}|\Psi_{K^-}(\vec{r})|^2 
 \Gamma_m(\vec{p}_1,\vec{p}_2,\vec{p}_K,\vec{r}),
\label{eq2a}
\end{equation}
where $|\Psi_{K^-}(\vec{r})|^2$ is the probability of finding 
the $K^-$ in the
nucleus, $|\vec{p}_1|, |\vec{p}_2| <k_F(r)$ with
$k_F(r)=\left( 3\pi^2 \rho(r)/2 \right)^{1/3}$ being
the local Fermi momentum and
$\Gamma_m(\vec{p}_1,\vec{p}_2,\vec{p}_K,\vec{r})$ is the in-medium
decay width for the $K^- p N \rightarrow \Lambda N$ process. The structure of
the integrals determining $\Gamma_m$,
\begin{equation}
\Gamma_m \propto \int d^3
\vec{p}_\Lambda d^3 \vec{p}_N \dots
K(\vec p_\Lambda,\vec r) K(\vec p_N,\vec r) \ ,
\end{equation}
allows us to follow 
the propagation of the produced nucleon and $\Lambda$ through the
nucleus after $K^-$ absorption via
the kernel $K(\vec p, \vec r)$. The former two equations describe
the process in which a kaon at rest is absorbed by two nucleons
($pp$ or $pn$) within the local Fermi sea, emitting a nucleon and
a $\Lambda$. The primary nucleon ($\Lambda$) is
 allowed to re-scatter with
nucleons in the nucleus according to a probability per unit length
given by $\sigma_{N(\Lambda)} \rho(r)$, where $\sigma_{N(\Lambda)}$ is the
experimental $NN(\Lambda N)$ cross section at the corresponding energy. We take $\sigma_{\Lambda}$ cross section fitted to the data from \cite{lambdan,lambdansigma0}:
\begin{equation}
\sigma_{\Lambda} =(39.66-100.45 x+92.44 x^2-21.40 x^3)/p_{LAB}\ {\rm [mb]}\,,
\label{sLN}
\end{equation}
where $x=Min(2.1$ GeV, $p_{LAB})$. 
In
\cite{magas06} a simpler parameterization for the $\Lambda$, of the type
$\sigma_\Lambda=2\sigma_N/3$ was employed, but as we shall see,  this modification 
affects non-negligibly only the results for heavy nuclei \cite{qnp06_proc}. 
 
We note that particles move under the
influence of a mean field potential, of Thomas-Fermi type.
The hole nucleon spectrum has an
additional constant binding $\Delta$ of a few MeV that forces
the maximum $\Lambda N$ invariant mass allowed
by our model, $m_{K^-} + 2 M_p -2\Delta$, to coincide with the
actual invariant mass reached in $K^-$ absorption leading to the
ground state of the daughter nucleus, $m_{K^-}+
M(A,Z)-M(A-2,Z-2)$. 
After several possible collisions, one or more
nucleons and a $\Lambda$ emerge from the nucleus and the invariant
mass of all possible $\Lambda p$ pairs, as well as their relative
angle, are evaluated for each Monte Carlo event.
See Ref.~\cite{magas06} for more details.

Absorption of a $K^-$ from a nucleus leaving the final daughter
nucleus in its ground state gives rise to a narrow peak in the
$\Lambda p$ invariant mass distribution, as observed in the
spectrum of \cite{finuda}. We note that our local density formalism, in which
the hole levels in the Fermi sea form a continuum of states,
cannot handle properly transitions to discrete states of the
daughter nucleus, in particular to the ground state.\footnote{
However including an
additional constant binding $\Delta$ for the hole nucleon spectrum we make sure that the 
position of the narrow peak for ground state to ground state transition is correct.} For this
reason, we will remove in our calculations those events in which
the $p$ and $\Lambda$ produced after $K^-$ absorption leave the
nucleus without having suffered a secondary collision. However,
due to the small overlap between the two-hole initial state after
$K^-$ absorption and the residual $(A-2)$ ground state of the
daughter nucleus, as well as to the limited
 survival probability for
both the $p$ and the $\Lambda$ crossing the nucleus without any collision,
this strength represents only a
moderate fraction, estimated to be smaller than 15\% in $^{7}$Li \cite{magas06}.\footnote{
It should also be mentioned here that the possibility of removing
two protons from the initial nucleus leading to a final
$(A-2)$ excited nucleus in the continuum without further interaction of the
$p$ or $\Lambda$ is negligible due to the small overlap between these two states.}

Our invariant mass spectra requiring at least one secondary
collision of the $p(n)$ or $\Lambda$ after the $K^- pp(np)$
absorption process are shown in Fig.~\ref{fig:1} for different
nuclei\footnote{The model presented here is based on Fermi sea approximation for original and daughter 
nuclei, and hence it is 
not valid for very small systems, like $K^-\ ^4$He.}, where we have
applied the same cuts as in the experiment, namely  $P_\Lambda >
300$ MeV/c (to eliminate events from $K^- p \rightarrow \Lambda
\pi$) and $\cos \Theta_{\vec{p}_\Lambda \vec{p}_p}<-0.8$ (to
filter $\Lambda p$ pairs going back-to-back). 
The maximum number of allowed secondary collisions  is 
2 for $^6$Li, 3 for $^{7}$Li, $^{9}$Be, $^{12}$C, $^{13}$C, 
$^{16}$O, 4 for $^{27}$Al and 5 for $^{51}$V.
The calculated
angular distribution shown in Ref.~\cite{magas06} demonstrates
that, even after collisions, a sizable fraction of the events
appear at the back-to-back kinematics. These events  generate
the main bump at 2260-2270 MeV 
in all the $\Lambda p$ invariant mass spectra shown in Fig.~\ref{fig:1}, at
about the same position as the main
peak shown in the inset of Fig. 3 of \cite{finuda}.
Since one
measures the $\Lambda p$ invariant mass, the main contribution
comes from $K^- p p \rightarrow \Lambda p$ absorption (dot-dashed
lines), although the contribution from the $K^- p n \rightarrow
\Lambda n$ reaction followed by $np\rightarrow pn$ (dotted line)
is non-negligible. 
It is interesting that the width
of the distribution
gets broader with the size of the nucleus, while the peak remains in the same
location, consistently to what one expects for the behavior of a quasi-elastic
peak. 
Let us point out in this context that the possible interpretation of the FINUDA
peak as a bound state of the $K^-$ with the nucleus,
not as a $K^- pp$ bound state,
would unavoidably lead to peaks at different energies for different nuclei \cite{Mares:2006vk}.
We finally observe
that the spectra of heavy nuclei develop a secondary peak at lower invariant masses
due to the larger amount of re-scattering processes. It is more
pronounced than that shown in our earlier work \cite{magas06}, due to the use
here of a realistic $\Lambda N$ scattering cross section. However, we want to stress 
that in the low invariant mass region other processes, not taken into account in our 
calculations, like for example one nucleon absorption, are important 
and may significantly modify the spectrum.

\begin{figure}
\begin{center}
\includegraphics[width=11.0cm,height=16.4cm,angle=-90]{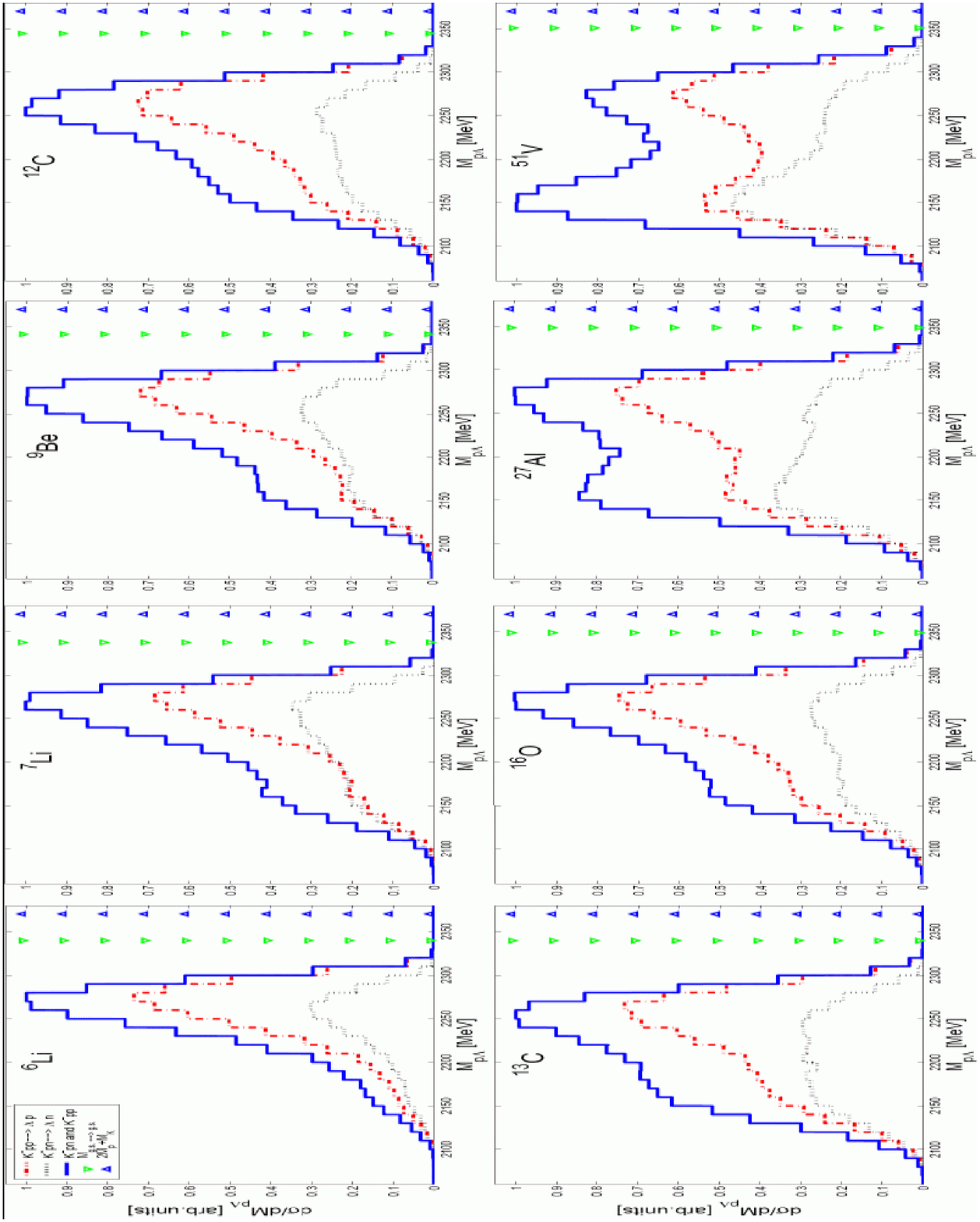}
\end{center}
\vspace{-0.5cm}
\caption{$\Lambda p$ invariant mass distribution for $K^-$ 
absorption in several nuclei imposing
 $P_\Lambda > 300$ MeV/c and 
$\cos \Theta_{\vec{p}_\Lambda \vec{p}_p}<-0.8$.
}
\vspace{-0.5cm}
\label{fig:1}       
\end{figure}

Finally, we compare our results with those presented in the inset
of Fig. 3 in Ref.~\cite{finuda}, using the three lighter targets
in the same proportion as in the experiment - see Fig. 2. We note that the averaged histogram
is dominated by the $^{12}$C component of the mixture, due mostly
to the larger overlap with the kaon wave function. As we see, our
calculations are in excellent agreement with the measured spectrum
in this range.

\begin{figure}
\begin{center}
\includegraphics[width=9.0cm]{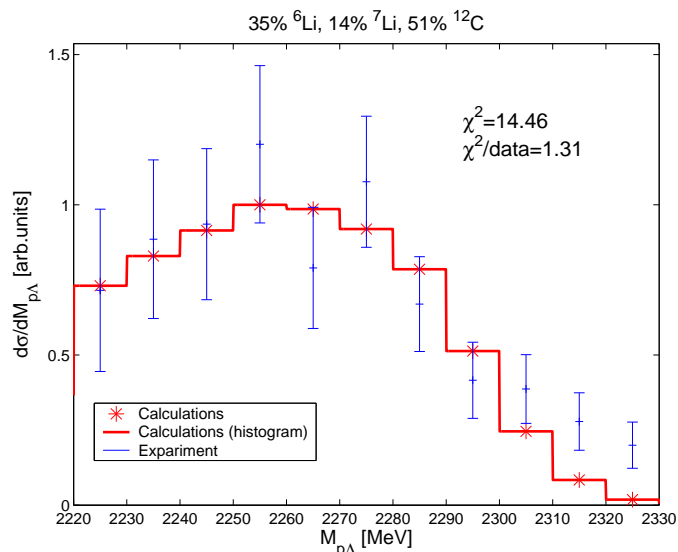}
\vspace{-0.5cm}
\caption{Invariant mass of $\Lambda p$ distribution
for $K^-$ absorption in light nuclei in the following proportion:
51\% $^{12}$C,  35\% $^{6}$Li and 14\% $^{7}$Li. Stars and
histogram show our results, while experimental points and
errorbars are taken from Ref. \cite{finuda}.}
\end{center}
\vspace{-1.0cm}
\label{chi2}
\end{figure}

Thus, the experimental $\Lambda p$
invariant mass spectrum of the FINUDA collaboration \cite{finuda}
is naturally explained in our Monte Carlo simulation as a
consequence of final state interactions of the particles produced
in nuclear $K^-$ absorption as they leave the nucleus, without the
need of resorting to exotic mechanisms like the formation of a
$K^- pp$ bound state.
Together with the interpretation of the proton momentum spectrum given in
Refs.~\cite{oset-toki,finuda_np06},
it seems then clear that there is at present no experimental evidence for the
existence of deeply bound kaonic states.

Recently new results came from few body calculations, either solving Faddeev 
equations \cite{nina_hyp06} or applying variational
techniques \cite{dote_hyp06}, using realistic $\bar{K}N$ interactions
and short-range nuclear correlations. These
works predict few-nucleon kaonic states bound by 50-70 MeV but having
large widths of the order of 100 MeV, thereby disclaiming the findings 
of Refs.~\cite{akaishi02,akaishi05}.

{\bf Acknowledgments. }\ This work is partly supported by contracts 
BFM2003-00856 and FIS2005-03142 from MEC (Spain) and FEDER,
the Generalitat de Catalunya contract 2005SGR-00343,
and the E.U. EURIDICE network contract HPRN-CT-2002-00311.
This research is part of the EU Integrated Infrastructure Initiative
Hadron Physics Project under contract number RII3-CT-2004-506078.

}

\end{document}